# Spatial non-locality in confined quantum systems: a liaison with quantum correlations

Ivan P. Christov

Institute of Electronics, Bulgarian Academy of Sciences, 1784 Sofia, Bulgaria

Physics Department, Sofia University, 1164 Sofia, Bulgaria

Email: ivan.christov@phys.uni-sofia.bg

ORCID ID: 0000-0002-9146-6708

**Abstract**

Using advanced stochastic methods (time-dependent quantum Monte Carlo, TDQMC) we explore the ground state of 1D and 2D artificial atoms with up to six bosons in harmonic trap where these interact by long-range and short-range Coulomb-like potentials (bosonic quantum dots). It is shown that the optimized value of the key variational parameter in TDQMC named nonlocal correlation length is close to the standard deviation of the Monte Carlo sample for one boson and it is slightly dependent on the range of the interaction potential. Also it is almost independent on the number of bosons for the 2D system thus confirming that the spatial quantum non-locality experienced by each particle is close to the spatial uncertainty exhibited by the rest of the particles. The intimate connection between spatial non-locality and quantum correlations is clearly evidenced.

## 1. Introduction

In classical physics, when measuring the position of a particle the measurement result corresponds to a certain location while in quantum mechanics the outcome is a whole set of possible locations. In the one-body case (e.g. in the hydrogen atom) the probability distribution of these locations is described by modulus square of the ground state wave function in coordinate representation which is of non-zero width because that state is not an eigenstate of the position operator [1]. In the many-body case, however, the wave function resides in configuration space which implies a mutual connection between the possible positions occupied by each particle with the positions of the rest of the particles, which evidences the quantum nonlocality and entanglement [2]. Using a Monte Carlo (MC) language one may assign an ensemble of finite number of point-like walkers to each physical particle where the standard deviation of the MC sample is a quantitative measure for how much uncertainty there is for that particle in coordinate space, as done in e.g. the variational quantum Monte Carlo (QMC) method, where a many-body trial wave function is sampled [3]. Other approaches to study few-particle quantum systems include Gaussian variational [4,5] , hyperspherical [6], and stochastic variational method [7,8] which also employ optimized static many-body trial functions. However the use of static trial functions may not be sufficient for describing some essential properties of quantum systems which depend on the phase of the wave function e.g. for real-time dynamics. Within the MC methodology one possible way to overcome the exponential scaling imposed by the exact many-body wave function is to introduce in a self-consistent manner an ensemble of wave-functions considered as random walkers in one-body Hilbert space for each particle, where each individual MC walker is guided by a corresponding wave-function (guide wave) in physical space and, at the same time, it samples the distribution given by the modulus square of that guide wave (particle-wave dichotomy). An efficient approach to achieve such particle-wave dichotomy is offered by the recent time-dependent quantum Monte Carlo (TDQMC) method [9]-[11] where random windowing applied to the Hartree-Fock approximation results in an ensemble of guide waves which obey a set of coupled one-body time-dependent Schrödinger equations (TDSE), where the interactions between the particles are accounted for using effective potentials with no self-interactions involved. A key ingredient in the TDQMC approach is that the effective potential is formulated in a form of Monte Carlo convolution which involves explicitly the spatial quantum nonlocality as a function of the variational parameter named nonlocal correlation length.

In this way, by reducing the many-body Schrödinger equation to a set of coupled one-body Schrödinger equations with stochastic potentials between the particles one can build an efficient time-dependent approach to treat quantum many-body systems which scales almost linearly with the system size and offers a reasonable tradeoff between scaling and accuracy. It is important to point out that since in TDQMC the walkers are guided through first-order de Broglie-Bohm equations, its predictions should not be related to Bohmian mechanics which uses quantum potentials.

In this paper the TDQMC method is used to show that for parabolic core potential the range of the spatial nonlocality experienced by bosons in 1D and 2D artificial atoms (bosonic quantum dots) is limited from below by the quantum uncertainty (standard deviation of the MC sample), for both long-range and short-range interaction potentials. We study also the linear entropy as an entanglement (and correlation) measure for the bosons in the trap. Other work on bosons in harmonic trap includes some exact analytical and numerical solutions for the Hooke model [12-14].

## 2. Methods

In the TDQMC method an ensemble of walkers and corresponding guide waves is attached to each physical particle such that the many-body Schrödinger equation is reduced to a number of one-body Schrödinger equations for the guide waves, which for the k-th walker from the i-th particle ensemble reads [10],[11]:

$$i\hbar \frac{\partial}{\partial t} \varphi_i^k(\mathbf{r}_i, t) = \left[ -\frac{\hbar^2}{2m_i} \nabla_i^2 + V_{e-n}(\mathbf{r}_i) + V_{eff}^k(\mathbf{r}_i, t) \right] \varphi_i^k(\mathbf{r}_i, t) \tag{1}$$

$i=1,...,N;$
$k=1,...,M,$

where $V_{e-n}(\mathbf{r}_i)$ is the classical core potential and the effective interaction potential $V_{eff}^k(\mathbf{r}_i, t)$ is given by a Monte Carlo convolution of the true potential $V_{e-e}[\mathbf{r}_i, \mathbf{r}_j]$ and the kernel function $K$ which incorporates the specific nonlocal quantum correlation length $\sigma_j$:

$$V_{eff}^{k}(\mathbf{r}_i,t) = \sum_{j \neq i}^{N} \frac{1}{Z_j^k} \sum_{l}^{M} V_{e-e}\left[\mathbf{r}_i, \mathbf{r}_j^l(t)\right] K\left[\mathbf{r}_j^l(t), \mathbf{r}_j^k(t), \sigma_j\right], \quad (2)$$

where:

$$K\left[\mathbf{r}_j, \mathbf{r}_j^k(t), \sigma_j\right] = \exp\left(-\frac{\left|\mathbf{r}_j - \mathbf{r}_j^k(t)\right|^2}{2\sigma_j\left(\mathbf{r}_j^k,t\right)^2}\right) \quad (3)$$

and:

$$Z_j^k = \sum_{l=1}^{M} K\left[\mathbf{r}_j^l(t), \mathbf{r}_j^k(t), \sigma_j\right] \quad (4)$$

is the weighting factor.

In this way the effective potential seen by the k-th guide wave for the i-th particle in Eq.(2) involves the interaction potentials due to a number of walkers which belong to the j-th particle which lie within the non-local length $\sigma_j\left(\mathbf{r}_j^k,t\right)$ around $\mathbf{r}_j$. The connection between the trajectories $\mathbf{r}_i^k(t)$ and the guide waves $\varphi_i^k(\mathbf{r}_i,t)$ is given by the de Broglie-Bohm guiding equations for real-time propagation:

$$\mathbf{v}_i^k(t) = \frac{\hbar}{m_i} \text{Im}\left[\frac{\nabla_i \varphi_i^k(\mathbf{r}_i,t)}{\varphi_i^k(\mathbf{r}_i,t)}\right]_{\mathbf{r}_i = \mathbf{r}_i^k(t)}, \quad (5)$$

and it is given by a drift-diffusion process for the ground-state preparation (imaginary-time propagation):

$$d\mathbf{r}_i^k(\tau) = \mathbf{v}_i^{Dk} d\tau + \mathbf{\eta}_i(\tau)\sqrt{\frac{\hbar}{m_i}} d\tau, \quad (6)$$

where:

$$\mathbf{v}_i^{Dk}(\tau) = \frac{\hbar}{m_i}\left[\frac{\nabla_i \varphi_i^k(\mathbf{r}_i,\tau)}{\varphi_i^k(\mathbf{r}_i,\tau)}\right]_{\mathbf{r}_i = \mathbf{r}_i^k(\tau)} \quad (7)$$

is the drift velocity, and $\boldsymbol{\eta}(\tau)$ is Markovian stochastic process whose amplitude tends to zero toward steady state as $\tau^{0.2}$.

The use of the nonlocal length $\sigma_j$ to quantify the spatial nonlocality in Eqs.(1)-(2) allows us to connect two limiting cases where for $\sigma_j^k \to 0$ the set of TDQMC equations (1) is transformed into a set of linear Schrödinger equations coupled with a local pair-wise interaction potential $V_{e-e}^{eff}(\mathbf{r}_i,t) = V_{e-e}\left[\mathbf{r}_i, \mathbf{r}_j^k(t)\right]$, and the opposite case where for $\sigma_j \to \infty$ all walkers for a given particle are guided by the same guide wave in Eq.(1) which recovers the mean-field (Hartree-Fock) approximation. Then the problem of finding the ground state of a many-body system is reduced to a variational procedure where the energy of the system:

$$E = \frac{1}{M}\sum_{k=1}^{M}\left[\sum_{i=1}^{N}\left[-\frac{\hbar^2}{2m_i}\frac{\nabla_i^2 \varphi_i^k(\mathbf{r}_i^k)}{\varphi_i^k(\mathbf{r}_i^k)} + V_{e-n}(\mathbf{r}_i^k)\right] + \sum_{i>j}^{N} V_{e-e}(\mathbf{r}_i^k - \mathbf{r}_j^k)\right]_{\substack{\mathbf{r}_i^k = \mathbf{r}_i^k(\tau) \\ \mathbf{r}_j^k = \mathbf{r}_j^k(\tau)}} \quad (8)$$

is to be minimized with respect to the nonlocal lengths $\sigma_j$, which gives the optimal configuration of walkers + guide waves which can be used next to calculate averages and coherences (reduced density matrices). On the algorithmic side, the preparation of the ground state of the quantum system involves initialization of Monte Carlo ensembles of walkers and guide waves, followed by their concurrent propagation in imaginary time toward steady state in the presence of random component in walker's motion, in accordance with Eqs.(1-7). For purely variational calculation (no branching involved) the drift term in Eq.(6) is to be neglected.

### 3. Results

In order to explore the role of the boson-boson interaction on the spatial quantum nonlocality here we calculate the ground state of 1D and 2D artificial atoms with parabolic core potential $V_{e-n}(\mathbf{r}_i) = r_i^2/2$ (bosonic quantum dots). Unlike in the standard quantum dots where there is fermionic exchange interaction which reduces the overlap between the wave functions for the equal-spin electrons here we have practically identical probability distributions for all bosons in the trap, similarly to the 1S state occupied by opposite-spin electrons in the two-electron atom. In order to explore the role of the potential range on the boson-boson correlation we have chosen Yukawa-type of soft-core repulsive potential:

$$V_{e-e}\left[\mathbf{r}_i, \mathbf{r}_j\right] = \frac{e^{-ar}}{\sqrt{r^2 + b^2}}, \tag{9}$$

which for $a \to 0$ transforms to the widely used "soft-core" Coulomb potential [15] which avoids the cusp at $r \equiv |\mathbf{r}_i - \mathbf{r}_j| \to 0$, as seen in Fig.1.

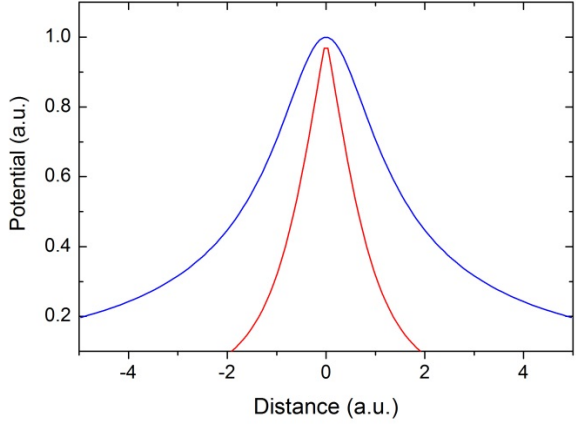

Fig.1. Long-range potential (a=0) –blue, and short-range potential (a=3) – red, with b=1 a.u. in Eq.(9).

For each walker from a given boson ensemble there is a number of corresponding walkers from the rest of ensembles where the spatial range of the interaction is characterized by the nonlocal length $\sigma_j$ in Eqs.(2)-(3). Intuitively, the nonlocal length is expected to be close to the spatial extend (uncertainty) of those ensembles which is given by their standard deviation $s_j$:

$$\sigma_j = \alpha_j \cdot s_j \qquad ; j=1,N \tag{10}$$

As a variational parameter, the nonlocal length $\sigma_j$ can be scanned either directly or through varying the parameter $\alpha_j$ in order to find the energy minimum. An impression on the size of the nonlocal length can be received from Fig.2(a) where the 2D configuration space is populated by the walkers of one-dimensional helium atom and the colored spot depicts the nonlocal range around a random walker pair for $\alpha_{1,2} = 1$ in Eq.(10), while Fig.2(b) shows the picture for $\alpha_{1,2} = 0.1$ which is close to the ultra-correlated case above ($\sigma_j \to 0$).

Through controlling the degree of correlation the spatial quantum non-locality controls also the quantum entanglement which is expected to be higher wherever the correlation is higher, which in our case approaches the ultra-correlated interaction, $\sigma_j \to 0$. In order to estimate the entanglement experienced by the i-th boson we can use its own reduced density matrix which can be straightforwardly calculated through the corresponding guiding waves [16]:

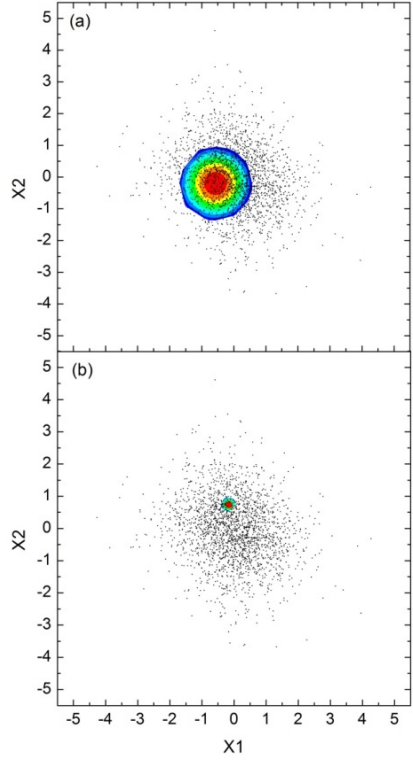

Fig.2. Spatial nonlocality in configuration space for two 1S electrons of 1D helium:
(a) - for nonlocal length $\sigma_j$ =1 a.u., (b) - for $\sigma_j$ =0.1 a.u..

$$\rho_i\left(\mathbf{r}_i,\mathbf{r}_i',t\right) = \frac{1}{M} \sum_{k=1}^{M} \varphi_i^{k*}(\mathbf{r}_i,t)\varphi_i^{k}(\mathbf{r}_i',t) \tag{11}$$

which can be used next to quantify the entanglement through e.g. the linear entropy:

$$S_L^i(t) = 1 - Tr(\rho_i^2) = 1 - \int \rho_i^2(\mathbf{r}_i, \mathbf{r}_i, t) d\mathbf{r}_i \qquad (12)$$

We have plotted in Fig.3 the energy (c) and the linear entropy (b) of 1D bosonic quantum dot with up to 6 particles, for long-range (a=0 in Eq.(9)) and short-range (a=3) repulsive potentials, with blue and with red lines, respectively. It is seen that within the statistical error there is a good correspondence between the TDQMC results and the exact numerical results (green lines) obtained by a direct solution of the many-body Schrödinger equation for up to 4 bosons in one spatial dimension.

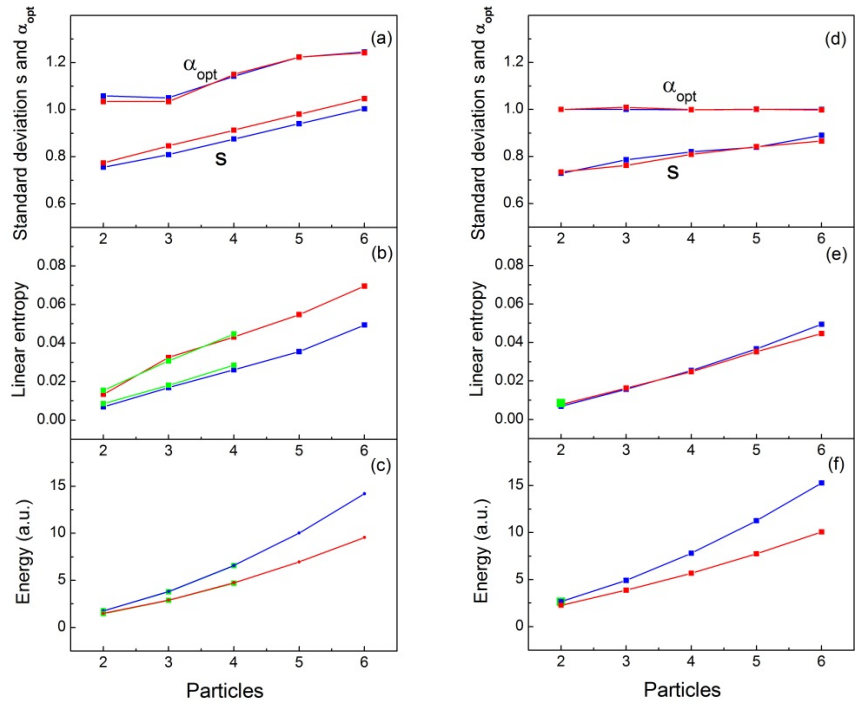

Fig.3. Spatial nonlocality parameter $\alpha_{opt}$ (a), (d), linear entropy (b), (e), and energy (c), (f) versus the number of bosons, for 1D quantum dot - (a) - (c) and for 2D quantum dot - (d) - (f). Blue lines for long-range potential, red lines for short-range potential, and green lines for numerically exact results.

The important result here is shown in Fig.3(a) where one can see that the parameter $\alpha_{opt} = \sigma_j / s_j$ which minimizes the energy of the system is almost independent on the range of the interaction potential despite the larger width of the probability distribution for the short-range potential which pushes the near-range bosons further from each other thus increasing the entropy and decreasing the overall energy of the system. Also it is seen that $\alpha_{opt}$ increases for up to 6 bosons while it is greater than one, which means that the nonlocal length $\sigma_j$ slightly exceeds the standard deviation of the MC sample in 1D. Note that since the interacting trapped bosons are identical we have dropped the subscript j in Fig.3.

These findings are further confirmed for two dimensional "quantum dots" where in Fig.3(d)-(f) the corresponding results are presented. Again, with increasing the number of particles the system energy is lower for the short-range potential while the linear entropy in Fig.3(e) only slightly varies for almost constant $\alpha_{opt}$ in Fig.3(d). The results of Fig.3 reflect the fact that for a given interaction potential the quantum correlations are stronger for lower spatial dimensions where the degrees of freedom of the particles are restricted. Although not surprising, the spatial nonlocality seen by each walker for a given boson depends on the spatial uncertainty exhibited by the other interacting bosons in accordance with the basic principles of quantum mechanics. Also, the effective independence of the parameter $\alpha_j$ on the range of the interaction potential suggests that at least for simple quantum systems at ground state the nonlocal length serves as a key parameter to control the degree of correlation and entanglement.

## 4. Conclusions

In conclusion, we have calculated the ground states of 1D and 2D model quantum dots with up to six boson, where we have focused on the influence of the spatial nonlocality on the correlations between the bosons. The results prove the important role played by the non-local length introduced within the time-dependent quantum Monte Carlo framework where the ratio $\alpha_j$ of the nonlocal length to the standard deviation which minimizes the ground state energy remains almost the same for long-range and short-range interaction potentials between the bosons. At the same time the linear entropy and the energy in the 1D case are close to the exact numerical results where these increase with the number of bosons. Because of the stronger correlations in reduced dimensionality, for the 1D model $\alpha_j$ still depends on the number of the interacting bosons while it is almost constant for the 2D model. Finally we should note that the intrinsic parallelism of the TDQMC method used here allowed us to calculate the ground state of the 1D quantum dot for less than a minute, with 20 000 walkers for each boson, and for a few minutes for the 2D case with up to 5 000 walkers.

**Acknowledgements:** This material is based upon work supported by the Air Force Office of Scientific Research under award number FA9550-19-1-7003.